\documentclass[conference]{IEEEtran}
\IEEEoverridecommandlockouts

\usepackage{cite, lipsum}
\usepackage{amsmath,amssymb,amsfonts}
\usepackage{algorithmic}
\usepackage{graphicx}
\usepackage{textcomp}
\usepackage{xcolor}
\usepackage{comment}
\usepackage{float, balance}
\usepackage{tikz}
\usepackage{msc}
\usepackage{algorithmic}
\usepackage{graphicx}
\usepackage{acronym, comment, subcaption}
\usepackage{tabularx, pifont}
\usepackage[ruled,vlined,linesnumbered]{algorithm2e}
\usepackage{hyperref}
\usepackage{mathtools}
\usepackage[colorinlistoftodos, textsize=small]{todonotes}
\usepackage{enumitem}
\usepackage{arydshln}
\usepackage{float}
\usepackage{flafter}
\usepackage{multirow}
\usepackage{booktabs}

\usepackage{tikz}
\usetikzlibrary{chains,shapes.multipart}
\usetikzlibrary{shapes,calc}
\usetikzlibrary{automata,positioning}

\SetCommentSty{mycommfont}

\newcommand\ahmedupdate[1]{\textcolor{black}{#1}}
\newcolumntype{P}[1]{>{\centering\arraybackslash}p{#1}}

\usetikzlibrary{calc,positioning,shapes,decorations.pathreplacing}

\tikzset{
short/.style={draw,rectangle,text height=3pt,text depth=13pt,
  text width=7pt,align=center,fill=gray!30},
long/.style={short,text width=1.5cm}
}

\usepackage[ruled,vlined]{algorithm2e}
\def\BibTeX{{\rm B\kern-.05em{\sc i\kern-.025em b}\kern-.08em
    T\kern-.1667em\lower.7ex\hbox{E}\kern-.125emX}}
\acrodef{AND}{Authenticated Neighbor Discovery}
\acrodef{AP}{Access Point}
\acrodef{AWGN}{Additive White Gaussian Noise}
\acrodef{CA}{Certificate Authority}
\acrodef{CI}{Confidence Interval}
\acrodef{CRCS}{Challenge Reflection with Channel Selection}
\acrodef{ECDSA}{Elliptic Curve Digital Signature Algorithm}
\acrodef{FHE}{Full Homomorphic Encryption}
\acrodef{HE}{Homomorphic Encryption}
\acrodef{HMAC}{Hash Message Authentication Code}
\acrodef{IoD}{Internet of Drones}
\acrodef{IoMT}{Internet of Medical Things}
\acrodef{IoT}{Internet of Things}
\acrodef{IoV}{Internet of Vehicles}
\acrodef{LBS}{Location-based Services}
\acrodef{LHE}{Linearly-Homomorphic Encryption}
\acrodef{LQ}{Link Quality}
\acrodef{LoS}{Line of Sight}
\acrodef{LTC}{Long Term Certificate}
\acrodef{LTCA}{Long Term Certification Authority}
\acrodef{MAC}{Medium Access Control}
\acrodef{MANETs}{Mobile Ad-hoc Networks}
\acrodef{MITM}{Man-In-The-Middle}
\acrodef{ND}{Neighbor Discovery}
\acrodef{NLoS}{Non-Line-of-Sight}
\acrodef{PCA}{Pseudonym Certificate Authority}
\acrodef{PE-SND}{Privacy-Enhancing \ac{SND}}
\acrodef{PHE}{Partially Homomorphic Encryption}
\acrodef{PKI}{Public Key Infrastructure}
\acrodef{PP-SND}{Privacy-Preserving Secure Neighbor Discovery}
\acrodef{PPLE}{Privacy-Preserving Location Estimation}
\acrodef{RF}{Radio Frequency}
\acrodef{RSS}{Radio Signal Strength}
\acrodef{RTT}{Round Trip Time}
\acrodef{SDR}{Software Defined Radio}
\acrodef{SHE}{Somewhat Homomorphic Encryption}
\acrodef{SMC}{Secure Multi-party Computation}
\acrodef{SND}{Secure Neighbor Discovery}
\acrodef{ToF}{Time of Flight}
\acrodef{UWB}{Ultra-Wide-Band}
\acrodef{VANET}{Vehicle Ad-hoc Network}

\graphicspath{{Figures/}}

\begin{document}

\title{Privacy-Preserving Secure Neighbor Discovery for Wireless Networks}

\author{
\IEEEauthorblockN{Ahmed Mohamed Hussain}
    \IEEEauthorblockA{
        Networked Systems Security Group\\
        \textit{KTH Royal Institute of Technology}\\ Stockholm, Sweden\\
        ahmed.hussain@ieee.org
    }

\and

\IEEEauthorblockN{Panos Papadimitratos}
    \IEEEauthorblockA{
        Networked Systems Security Group\\
        \textit{KTH Royal Institute of Technology}\\ Stockholm, Sweden\\
        papadim@kth.se
    }
}

\maketitle

\begin{abstract}
	Traditional \acf{ND} and \acf[4]{SND} are key elements for network functionality. \ac{SND} is a hard problem, satisfying not only typical security properties (authentication, integrity) but also verification of direct communication, which involves distance estimation based on time measurements and device coordinates. Defeating relay attacks, also known as ``wormholes'', leading to stealthy Byzantine links and significant degradation of communication and adversarial control, is key in many wireless networked systems. However, \ac{SND} is not concerned with privacy; it necessitates revealing the identity and location of the device(s) participating in the protocol execution. This can be a deterrent for deployment, especially involving user-held devices in the emerging \ac{IoT} enabled smart environments. To address this challenge, we present a novel \acf{PP-SND} protocol, enabling devices to perform \ac{SND} without revealing their actual identities and locations, effectively decoupling discovery from the exposure of sensitive information. We use \acf{HE} for computing device distances without revealing their actual coordinates, as well as employing a pseudonymous device authentication to hide identities while preserving communication integrity. \ac{PP-SND} provides \ac{SND}~\cite{poturalski2013formal} along with pseudonymity, confidentiality, and unlinkability. Our presentation here is not specific to one wireless technology, and we assess the performance of the protocols (cryptographic overhead) on a Raspberry Pi 4 and provide a security and privacy analysis.
\end{abstract}


\section{Introduction}
\label{sec:intro}

Mobile networks are integral to our daily lives, offering essential services through various applications, building on the growth of 5G technologies~\cite{5gchanges}, and the \ac{IoT}, or more specifically, \ac{IoV}, the \ac{IoMT}, or the \ac{IoD}. These networks need security, but they also affect the privacy of the users they interact with. A critical aspect of these technologies is \acf{ND}, enabling devices to discover and communicate with other nearby devices~\cite{papadimitratos2008secure}. \ac{ND} protocols, however, are susceptible to eavesdropping, spoofing, \ac{MITM} attacks, and relay or wormhole attacks~\cite{alsa2012secure, papadimitratos2008secure}. \acf{SND} addresses these vulnerabilities with the use of cryptographic primitives and device distance estimation based on time and/or location measurements~\cite{papadimitratos2008secure}. Essentially, two nodes executing an \ac{SND} protocol measure their distance in two ways, including exchanging their coordinates and authenticating each other. 

Although \ac{SND} can provably defeat attacks, notably including relays, curious observers learn which device is a neighbor of which and the locations of nearby devices that run \ac{SND}. Even if neighbor-to-neighbor encryption were applied, curious yet honest peers running themselves the \ac{SND} protocol could learn the location of their peers. 

The challenge is to both have \ac{SND} and prevent the disclosure of identity and location information. Existing \ac{SND} solutions are not concerned with privacy dimension~\cite{papadimitratos2008secure}. Existing security solutions, IPsec, TLS, (but also SEND, WPA2/3, and cellular 4/5G LTE security) do not solve the \ac{SND} problem – they all provide authentication, integrity, some access control, and confidentiality. In other words, they provide only \ac{AND} with a level of privacy protection but not \ac{SND}. 

Integrating an \ac{SND} protocol with existing security protocols that offer confidentiality is not easy and would duplicate or, at best, complex functionality with pairwise security associations (e.g., an IPsec device to device tunnel and AH and ESP) and run the \ac{SND} protocol with them. On the other hand, AND can be combined with privacy-enhancing approaches, e.g., by changing/randomizing the MAC and IP address of the device each time it connects to an \ac{AP}~\cite{hugon2022roma, martin2017study} or more recent proposals that re-randomize addresses per packet~\cite{jin2024over}. None of these approaches are concerned with \ac{SND}. Nonetheless, the notion of ephemeral and changing identities can be useful. Overall, special care is needed in any composition of security protocols to ensure the sought properties are satisfied---in particular, \ac{SND} correctness and availability---after the privacy-enhancing/preserving elements are added. Therefore, focusing on \ac{SND} solutions, identities and locations are left vulnerable during the \ac{SND} process. 

We close this gap by introducing the first \acf{PP-SND} protocol to ensure secure and private node discovery for next-generation networks. \acf{HE} conceals data as it allows performing computations with encrypted data~\cite{acar2018survey}. \ac{HE}, together with pseudonymous authentication, are the building blocks for \ac{PP-SND}. 

The main objective of our protocol is to prevent learning about the identity of the nodes participating in the protocol execution and their locations. This ensures (i) secure and pseudonymous \ac{SND}, specifically addressing the challenge of an \textit{honest-but-curious} adversary that executes the \ac{SND} protocol with the intent to obtain sensitive data (location and identity) of other devices/users. (ii) effectively preventing internal and external adversaries from being able to identify, extract, or link information eavesdropped during the protocol execution by any of the devices in their neighborhood.

\usetikzlibrary{decorations.pathmorphing}
\tikzset{every picture/.style={line width=0.75pt}}
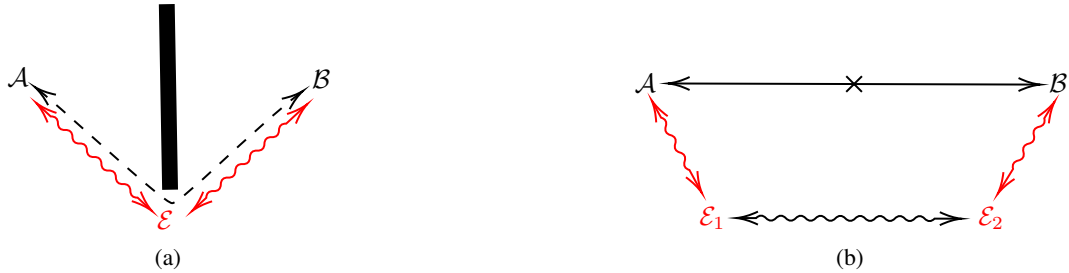
\begin{figure*}[!htbp]
\vspace{-0.25cm}
\centering
\begin{minipage}[b]{0.5\textwidth}
    \centering
    \tikzset{every picture/.style={line width=0.75pt}}

    \begin{tikzpicture}[x=0.75pt,y=0.75pt,yscale=-1,xscale=1]
    \draw [line width=6]    (176.28,130.22) -- (177.95,223.89) ;
    \draw  [dash pattern={on 4.5pt off 4.5pt}]  (111.44,173.38) .. controls (127.03,185.82) and (174.52,230.67) .. (179,230.67) .. controls (183.57,230.67) and (218.88,194.17) .. (243.51,175.14) ;
    \draw [shift={(245,174)}, rotate = 142.88] [color={rgb, 255:red, 0; green, 0; blue, 0 }  ][line width=0.75]    (10.93,-3.29) .. controls (6.95,-1.4) and (3.31,-0.3) .. (0,0) .. controls (3.31,0.3) and (6.95,1.4) .. (10.93,3.29)   ;
    \draw [shift={(109.67,172)}, rotate = 36.87] [color={rgb, 255:red, 0; green, 0; blue, 0 }  ][line width=0.75]    (10.93,-3.29) .. controls (6.95,-1.4) and (3.31,-0.3) .. (0,0) .. controls (3.31,0.3) and (6.95,1.4) .. (10.93,3.29)   ;
    
    \draw (94.43,161.13) node [anchor=north west][inner sep=0.75pt]   [align=left] {$\mathcal{A}$};
    \draw (248.18,161.76) node [anchor=north west][inner sep=0.75pt]   [align=left] {$\mathcal{B}$};
    \draw (169.89,233.05) node [anchor=north west][inner sep=0.75pt]   [align=left] {\color{red}$\mathcal{E}$};
    \draw [color={rgb, 255:red, 255; green, 0; blue, 4 }  ,draw opacity=1 ]   (111.89,181.44) -- (117.74,186.9) .. controls (120.09,186.83) and (121.31,187.97) .. (121.39,190.32) .. controls (121.47,192.67) and (122.69,193.81) .. (125.04,193.73) .. controls (127.39,193.65) and (128.61,194.79) .. (128.7,197.14) .. controls (128.78,199.49) and (130,200.63) .. (132.35,200.56) .. controls (134.7,200.48) and (135.92,201.62) .. (136,203.97) .. controls (136.08,206.33) and (137.3,207.47) .. (139.66,207.39) .. controls (142.01,207.31) and (143.23,208.45) .. (143.31,210.8) .. controls (143.39,213.15) and (144.61,214.29) .. (146.96,214.21) .. controls (149.32,214.13) and (150.54,215.27) .. (150.62,217.63) .. controls (150.7,219.98) and (151.92,221.12) .. (154.27,221.04) .. controls (156.62,220.97) and (157.84,222.11) .. (157.92,224.46) .. controls (158,226.81) and (159.22,227.95) .. (161.57,227.87) -- (161.58,227.88) -- (167.42,233.34) ;
    \draw [shift={(168.89,234.7)}, rotate = 223.06] [color={rgb, 255:red, 255; green, 0; blue, 4 }  ,draw opacity=1 ][line width=0.75]    (10.93,-3.29) .. controls (6.95,-1.4) and (3.31,-0.3) .. (0,0) .. controls (3.31,0.3) and (6.95,1.4) .. (10.93,3.29)   ;
    \draw [shift={(110.43,180.07)}, rotate = 43.06] [color={rgb, 255:red, 255; green, 0; blue, 4 }  ,draw opacity=1 ][line width=0.75]    (10.93,-3.29) .. controls (6.95,-1.4) and (3.31,-0.3) .. (0,0) .. controls (3.31,0.3) and (6.95,1.4) .. (10.93,3.29)   ;
    \draw [color={rgb, 255:red, 255; green, 0; blue, 0 }  ,draw opacity=1 ]   (190.35,233.41) -- (196.21,227.96) .. controls (196.3,225.61) and (197.53,224.47) .. (199.88,224.56) .. controls (202.23,224.65) and (203.45,223.51) .. (203.54,221.16) .. controls (203.63,218.81) and (204.86,217.67) .. (207.21,217.76) .. controls (209.56,217.85) and (210.78,216.71) .. (210.87,214.36) .. controls (210.96,212) and (212.18,210.86) .. (214.54,210.95) .. controls (216.89,211.04) and (218.11,209.9) .. (218.2,207.55) .. controls (218.29,205.2) and (219.52,204.06) .. (221.87,204.15) .. controls (224.22,204.24) and (225.44,203.1) .. (225.53,200.75) .. controls (225.62,198.4) and (226.84,197.26) .. (229.19,197.35) .. controls (231.54,197.44) and (232.77,196.3) .. (232.86,193.95) .. controls (232.95,191.6) and (234.17,190.46) .. (236.52,190.54) -- (239.85,187.45) -- (245.72,182.01) ;
    \draw [shift={(247.18,180.65)}, rotate = 137.13] [color={rgb, 255:red, 255; green, 0; blue, 0 }  ,draw opacity=1 ][line width=0.75]    (10.93,-3.29) .. controls (6.95,-1.4) and (3.31,-0.3) .. (0,0) .. controls (3.31,0.3) and (6.95,1.4) .. (10.93,3.29)   ;
    \draw [shift={(188.89,234.77)}, rotate = 317.13] [color={rgb, 255:red, 255; green, 0; blue, 0 }  ,draw opacity=1 ][line width=0.75]    (10.93,-3.29) .. controls (6.95,-1.4) and (3.31,-0.3) .. (0,0) .. controls (3.31,0.3) and (6.95,1.4) .. (10.93,3.29)   ;
\end{tikzpicture}\\
\subcaption{}
\label{subfig:internal_adv}
\end{minipage}%
\begin{minipage}[b]{0.5\textwidth}
    \centering
    \begin{tikzpicture}[x=0.75pt,y=0.75pt,yscale=-1,xscale=1]
        \draw (352,109) node [anchor=north west][inner sep=0.75pt]  [align=left] {$\mathcal{A}$};
        \draw (560.98,109) node [anchor=north west][inner sep=0.75pt]  [align=left] {$\mathcal{B}$};
        \draw (384.69,175.97) node [anchor=north west][inner sep=0.75pt]  [align=left] {\color{red} $\mathcal{E}_1$};
        \draw (525.29,175.97) node [anchor=north west][inner sep=0.75pt]  [align=left] {\color{red} $\mathcal{E}_2$};
        \draw [color={rgb, 255:red, 255; green, 0; blue, 0 }  ,draw opacity=1 ]   (362.96,125.41) -- (366.77,132.44) .. controls (369.03,133.11) and (369.82,134.58) .. (369.15,136.84) .. controls (368.48,139.1) and (369.27,140.57) .. (371.53,141.24) .. controls (373.79,141.91) and (374.58,143.37) .. (373.91,145.63) .. controls (373.24,147.89) and (374.03,149.36) .. (376.29,150.03) .. controls (378.55,150.7) and (379.34,152.16) .. (378.67,154.42) .. controls (378,156.68) and (378.79,158.15) .. (381.05,158.82) -- (382.65,161.77) -- (386.46,168.81) ;
        \draw [shift={(387.42,170.57)}, rotate = 241.56] [color={rgb, 255:red, 255; green, 0; blue, 0 }  ,draw opacity=1 ][line width=0.75]    (10.93,-3.29) .. controls (6.95,-1.4) and (3.31,-0.3) .. (0,0) .. controls (3.31,0.3) and (6.95,1.4) .. (10.93,3.29)   ;
        \draw [shift={(362,123.65)}, rotate = 61.56] [color={rgb, 255:red, 255; green, 0; blue, 0 }  ,draw opacity=1 ][line width=0.75]    (10.93,-3.29) .. controls (6.95,-1.4) and (3.31,-0.3) .. (0,0) .. controls (3.31,0.3) and (6.95,1.4) .. (10.93,3.29)   ;
        \draw [color={rgb, 255:red, 255; green, 0; blue, 0 }  ,draw opacity=1 ]   (564.23,125.98) -- (560.03,132.79) .. controls (560.58,135.08) and (559.7,136.5) .. (557.41,137.04) .. controls (555.12,137.59) and (554.24,139.01) .. (554.78,141.3) .. controls (555.33,143.59) and (554.45,145.01) .. (552.16,145.56) .. controls (549.87,146.1) and (548.99,147.52) .. (549.53,149.81) .. controls (550.08,152.1) and (549.2,153.52) .. (546.91,154.07) .. controls (544.62,154.62) and (543.74,156.04) .. (544.29,158.33) -- (541.74,162.46) -- (537.54,169.27) ;
        \draw [shift={(536.49,170.97)}, rotate = 301.65] [color={rgb, 255:red, 255; green, 0; blue, 0 }  ,draw opacity=1 ][line width=0.75]    (10.93,-3.29) .. controls (6.95,-1.4) and (3.31,-0.3) .. (0,0) .. controls (3.31,0.3) and (6.95,1.4) .. (10.93,3.29)   ;
        \draw [shift={(565.28,124.28)}, rotate = 121.65] [color={rgb, 255:red, 255; green, 0; blue, 0 }  ,draw opacity=1 ][line width=0.75]    (10.93,-3.29) .. controls (6.95,-1.4) and (3.31,-0.3) .. (0,0) .. controls (3.31,0.3) and (6.95,1.4) .. (10.93,3.29)   ;
        \draw    (406.69,183.1) -- (414.69,183.13) .. controls (416.36,181.46) and (418.02,181.47) .. (419.69,183.14) .. controls (421.35,184.81) and (423.02,184.82) .. (424.69,183.16) .. controls (426.36,181.5) and (428.02,181.5) .. (429.69,183.17) .. controls (431.35,184.84) and (433.02,184.85) .. (434.69,183.19) .. controls (436.36,181.53) and (438.02,181.53) .. (439.69,183.2) .. controls (441.35,184.87) and (443.02,184.88) .. (444.69,183.22) .. controls (446.36,181.56) and (448.02,181.56) .. (449.69,183.23) .. controls (451.35,184.9) and (453.02,184.91) .. (454.69,183.25) .. controls (456.36,181.59) and (458.02,181.59) .. (459.69,183.26) .. controls (461.35,184.93) and (463.02,184.94) .. (464.69,183.28) .. controls (466.36,181.62) and (468.02,181.62) .. (469.69,183.29) .. controls (471.35,184.96) and (473.02,184.97) .. (474.69,183.31) .. controls (476.36,181.65) and (478.02,181.65) .. (479.69,183.32) .. controls (481.34,184.99) and (483.01,185) .. (484.68,183.34) .. controls (486.35,181.68) and (488.01,181.68) .. (489.68,183.35) .. controls (491.34,185.02) and (493.01,185.03) .. (494.68,183.37) .. controls (496.35,181.71) and (498.01,181.71) .. (499.68,183.38) .. controls (501.34,185.05) and (503.01,185.06) .. (504.68,183.4) -- (508.29,183.41) -- (516.29,183.43) ;
        \draw [shift={(518.29,183.44)}, rotate = 180.17] [color={rgb, 255:red, 0; green, 0; blue, 0 }  ][line width=0.75]    (10.93,-3.29) .. controls (6.95,-1.4) and (3.31,-0.3) .. (0,0) .. controls (3.31,0.3) and (6.95,1.4) .. (10.93,3.29)   ;
        \draw [shift={(404.69,183.1)}, rotate = 0.17] [color={rgb, 255:red, 0; green, 0; blue, 0 }  ][line width=0.75]    (10.93,-3.29) .. controls (6.95,-1.4) and (3.31,-0.3) .. (0,0) .. controls (3.31,0.3) and (6.95,1.4) .. (10.93,3.29)   ;
        \draw    (370.23,115.18) -- (556.98,115.74) ;
        \draw [shift={(557.98, 115.75)}, rotate = 180.16] [color={rgb, 255:red, 0; green, 0; blue, 0 }  ][line width=0.75]    (10.93,-3.29) .. controls (6.95,-1.4) and (3.31,-0.3) .. (0,0) .. controls (3.31,0.3) and (6.95,1.4) .. (10.93,3.29)   ;
        \draw [shift={(464.11, 115.46)}, rotate = 45.16] [color={rgb, 255:red, 0; green, 0; blue, 0 }  ][line width=0.75]    (-5.59,0) -- (5.59,0)(0,5.59) -- (0,-5.59)   ;
        \draw [shift={(370.23, 115.17)}, rotate = 0.16] [color={rgb, 255:red, 0; green, 0; blue, 0 }  ][line width=0.75]    (10.93,-3.29) .. controls (6.95,-1.4) and (3.31,-0.3) .. (0,0) .. controls (3.31,0.3) and (6.95,1.4) .. (10.93,3.29)   ;
    \end{tikzpicture}\\
\subcaption{}
\label{subfig:external_adv}
\end{minipage}%
\caption{$\mathcal{A}$ and $\mathcal{B}$ are unable to discover one another due to (a) a challenging propagation environment, an obstacle, or (b) being far from each other. The adversary $\mathcal{E}$ is a relay with one or two devices, relaying messages between $\mathcal{A}$ and $\mathcal{B}$ and misleading them that they are neighbors while they are not.}
\end{figure*}

\noindent \textbf{Contribution.} In this paper, we identify and present the requirements for \ac{SND} to be privacy-preserving. Then, we introduce the first \ac{PP-SND} protocol that conceals the device identity and location when performing \ac{SND}. Additionally, we evaluate the \ac{PP-SND} protocol performance overhead and provide security \& privacy analysis to demonstrate feasibility and robustness.

\noindent{\bf Paper Organization.} Sec.~\ref{sec:tp_snd} discusses the classical Two-Party \ac{SND}, the adversary model, and variants of protocols solving the problem. Sec.~\ref{sec:adversarial} discusses the adversarial model relevant to privacy. Sec.~\ref{sec:problem_statement} presents the problem statement and the properties required for \ac{PP-SND}. Sec.~\ref{sec:protocol_design} presents the \ac{PP-SND} protocol. In Sec.~\ref{sec:performanceANDanalysis}, we provide a performance evaluation and the protocol security \& privacy analysis. Sec.~\ref{sec:rw} discusses relevant related literature, before we conclude and discuss future work in Sec.~\ref{sec:conclusion}.

\begin{table}[H]
\centering
\caption{Notations used throughout the paper.}
\resizebox{\columnwidth}{!}{%
\begin{tabular}{|c|l|}
\hline
\textbf{Symbol} & \textbf{Description} \\ \hline
$\mathcal{A}$, $\mathcal{B}$, $\mathcal{E}$   & Alice,  Bob, and Eve                \\ \hline
ToF    &       \acl{ToF}               \\ \hline
$d_{ToF}()$  &  Distance based on \acsfont{ToF}           \\ \hline
$d_{\text{loc}}()$  &  Distance based on location           \\ \hline
$d_{HE}()$  &  Distance based on \acf{HE} computation \\\hline
$\epsilon$ & Distance error threshold \\\hline
$cert$  &  Certificate\\\hline
$PNYM$  &  Pseudonym\\\hline
$\Delta$  &  Processing time\\\hline
$ppk$ & Paillier public key\\\hline
$psk$ & Paillier secret (private) key\\\hline
$\mathcal{R}_{SND}$ & \acl{SND} range\\\hline
$x$, $y$ & latitude, longitude  \\\hline
$X$, $Y$& Encrypted latitude, Encrypted longitude \\\hline

\end{tabular} %
}
\end{table}

\section{Two-Party Secure Neighbor Discovery}
\label{sec:tp_snd}
\textbf{\acf{ND}} is a fundamental network protocol mechanism that enables devices to identify each other's presence/proximity~\cite{papadimitratos2008secure}. Here, we focus on communication neighborhood, that is, the ability to communicate directly, essential for initializing and maintaining local network communications. \ac{ND} mechanisms typically rely on broadcasting or multicasting discovery messages that may include the device's \ac{MAC} address and possibly other identifiers (e.g., IP address) and its attributes.

\textbf{\acf{AND}} introduces cryptographic mechanisms for securing \ac{ND} messages. Digital signatures or keyed-\acp{HMAC} ensure that messages originate from the sending node and have not been tampered with in transit. The SEcure Neighbor Discovery (SEND), specified in RFC 3971~\cite{rfc3971}, utilizes Cryptographically Generated Addresses (CGAs) and a \ac{PKI} to authenticate the sender and protect the integrity of the messages.

\ac{AND} can thwart device impersonation and \ac{ND} forgery/modification attacks. However, it cannot alone ensure that two devices that exchanged authenticated messages, whose integrity and freshness can be verified, are indeed neighbors, that is, can directly communicate with each other. To clarify this, consider a scenario with two legitimate, honest devices, Alice ($\mathcal{A}$) and Bob ($\mathcal{B}$), attempting to establish a neighbor relationship through an authenticated ND protocol. In  Fig.~\ref{subfig:internal_adv}, $\mathcal{A}$ and $\mathcal{B}$ are not communication neighbors but an adversary, Eve ($\mathcal{E}$), positioned physically between $\mathcal{A}$ and $\mathcal{B}$, captures the authenticated discovery message from $\mathcal{A}$ and relays it to $\mathcal{B}$, and vice versa, without altering the content. Despite the messages being authenticated and their integrity and freshness being verifiable, $\mathcal{E}$'s intervention falsely convinces $\mathcal{A}$ and $\mathcal{B}$ of their direct neighbor relationship, as the protocol lacks the means to verify the physical source of the communication.

Extending the aforementioned simple relay attack, a remote relay attack involves more sophisticated adversarial capabilities, potentially leveraging a network of malicious nodes. In this scenario, in Fig.~\ref{subfig:external_adv}, $\mathcal{A}$ and $\mathcal{B}$ are located in distant network segments, far beyond each other's direct communication range, i.e., not within the communication neighborhood. The adversary, controlling a set of nodes $\{\mathcal{E}_1, \mathcal{E}_2\}$, creates a relay chain that captures the discovery message from $\mathcal{A}$ and forwards it through the other malicious node to $\mathcal{B}$, and vice versa. The relayed messages integrity and authenticity are maintained, misleading $\mathcal{A}$ and $\mathcal{B}$ into believing they are neighbors, i.e., capable of direct communication.

\textbf{\acf{SND}~\cite{papadimitratos2008secure, PapadimitratosH_2003}} extends \emph{Authenticated} \ac{ND} by combining cryptography and distance estimates based on \ac{ToF} measurements and geographical coordinates. This way, \ac{SND} authenticates nodes executing the protocol, ensuring the integrity (and possibly the confidentiality, if such a feature were added) of the discovery message(s), and can establish whether nodes are in direct communication, hence, protecting against relay attacks. A formally proven secure \ac{SND} that utilizes time or location-based distance estimate, or both combined for distance estimation, is presented in~\cite{poturalski2013formal}. The authors considered the adversarial model (depicted in Fig.~\ref{subfig:internal_adv} and~\ref{subfig:external_adv}) permits an external adversary to perform message relaying with a minimum relaying delay constraint, $\Delta_{\text{relay}}$.

\begin{figure}
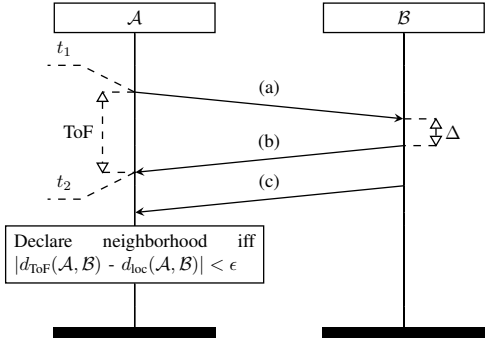

\vspace{-1cm}
\setlength{\envinstdist}{0cm}
    \centering
    \setmsckeyword{}
    \resizebox{0.75\columnwidth}{!}{%
    \begin{msc}[l]{}
        \setlength{\envinstdist}{2\envinstdist}
        \setlength{\instwidth}{3\mscunit}
        \setlength{\instdist}{1\mscunit} 
        \setlength{\instdist}{2cm}
        \declinst{A}{}{$\mathcal{A}$}
        \declinst{B}{}{$\mathcal{B}$}
        \nextlevel[.5]
        \mscmark{$t_{1}$}{A}
        \measure[side=left]{ToF}{A}{A}[3]
        \mess{(a)}{A}{B}[1]
        \nextlevel[1]
        \measure[side=right]{$\Delta$}{B}{B}[1]
        \nextlevel[1]
        \mess{(b)}{B}{A}[1]
        \nextlevel[1]
        \mscmark[position=below]{$t_{2}$}{A}
        \nextlevel[0.5]
        \mess{(c)}{B}{A}[1]
        \nextlevel[1.5]
        \action*{\parbox{4.5cm}{Declare neighborhood iff $|d_{\text{ToF}}(\mathcal{A}, \mathcal{B})$ - $d_{\text{loc}}(\mathcal{A}, \mathcal{B})| < \epsilon$ }}{A}
        \nextlevel[3]
        \end{msc}%
    }
    \caption{\textit{CR}-TL \ac{SND} Protocol variant. (a) is $\mathcal{A}$'s challenge $\langle$time, n$_{1}$$\rangle$, (b) is $\mathcal{B}$'s response $\langle$time, n$_{2}$$\rangle$, and (c) contains $\langle$time, location($\mathcal{B}$), auth$_{\mathcal{B}}$(n$_{1}$, n$_{2}$, location($\mathcal{B}$))$\rangle$. This variant does not require clocks to be synchronized.}
    \label{fig:TLCR}
    \vspace{-0.5cm}
\end{figure}

In a typical Two-Party \ac{SND} setup, two honest nodes, $\mathcal{A}$ and $\mathcal{B}$ in a wireless network, provisioned with cryptographic credentials, are legitimate network participants. In other words, the \ac{SND} protocol does not consider internal adversarial nodes (with credentials to participate yet deviate from the protocol operation). These nodes can be part of a larger network infrastructure where secure communication is essential, and they could be stationary or mobile, depending on the type of network topology. The \ac{SND} ensures that a node can discover its communication neighbors, even in the presence of adversaries. Two classes of proven secure \ac{SND} protocols exist: Time-based (T) protocols and Time-and-Location-based (TL) protocols~\cite{poturalski2013formal}, operating in a Beaconing (B) or Challenge-Response (CR) mode. Fig.~\ref{fig:TLB} and~\ref{fig:TLCR} illustrate the sequence of messages and operations for B and CR-TL  protocols.  

\textbf{Time-based \ac{SND} protocols} rely on the precise measurement of message propagation time, converting the time a message takes to travel between two devices into a distance estimate. B-T protocols require synchronized clocks among devices; if the measured distance does not exceed a predefined threshold, devices are within direct communication range. However, T protocols guarantee \ac{SND} if the relay adversary, $\Delta_{\text{relay}}$ is above a particular value~\cite{poturalski2013formal}.

\textbf{Time-and-Location-based \ac{SND} protocols} incorporate both precise timing and geographical location information, combining the device geographical distance with the distance estimated based on \ac{ToF}, offering a stronger basis for verifying neighbor relationships. Specifically, TL protocols compare the \ac{ToF} estimated distance with the geographical distance obtained from secure and accurate location information, achieving \ac{SND} for any $\Delta_{\text{relay}} > 0$~\cite{poturalski2013formal}.

\textbf{\textit{B} and \textit{CR} Variants} are variants T and TL protocols. Beacon-based (\textit{B}) protocols are designed for efficiency, requiring only a single message for \ac{SND} while participants maintain synchronized clocks to estimate the \ac{ToF}. The \textit{B} protocols are useful in scenarios where minimal communication overhead is desired. On the other hand, Challenge-Response (\textit{CR}) protocols involve a two-message exchange (a challenge followed by a response) for ranging (distance estimation). Unlike \textit{B} protocols, \textit{CR} ones do not require participants' synchronized clocks for \ac{ToF} estimation.

\subsection{Specifics and Comparison of Variants}
T protocols, relying solely on timing information to estimate distance, can be deceived by fast relaying of messages, misleading a node into believing it is closer than it actually is. TL protocols incorporate both timing and precise geographical location information to verify the proximity of the nodes. By comparing the measured \ac{ToF} of the signals with the calculated distance based on known locations, TL protocols can accurately determine whether the responding node is within the expected range. This dual verification makes TL protocols resilient against sophisticated, low-delay relay attacks, unlike T protocols that are effective only when the $\Delta_{\text{relay}}$ is above a certain threshold. 

\drawframe{no}
\begin{figure}
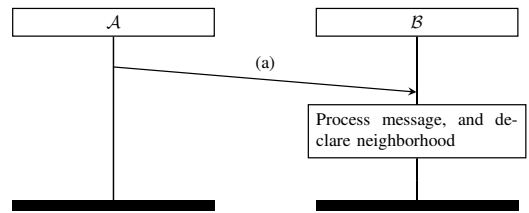

\vspace{-1.25cm}
\setlength{\envinstdist}{0cm}
    \centering
    \setmsckeyword{}
    \resizebox{0.8\columnwidth}{!}{%
    \begin{msc}[l]{}
        \setlength{\envinstdist}{2\envinstdist}
        \setlength{\instwidth}{4\mscunit}
        \setlength{\instdist}{1\mscunit} 
        \setlength{\instdist}{2cm}
        \declinst{A}{}{$\mathcal{A}$}
        \declinst{B}{}{$\mathcal{B}$}
        \mess{(a)}{A}{B}[1]
        \nextlevel[1.5]
        \action*{\parbox{4cm}{Process message, and declare neighborhood}}{B}
        \nextlevel[3]
        \end{msc}%
    }
    \caption{TL Beacon-based \ac{SND} protocol variant. Where message (a) is Beacon Message contains $\langle$time, location$\rangle$. This variant requires both participants to have synchronized clocks.}
    \label{fig:TLB}
\end{figure}

\drawframe{no}

For the \textit{B}-TL (Fig.~\ref{fig:TLB}) \ac{SND} protocol, $\mathcal{A}$ sends a beacon message received by $\mathcal{B}$. The beacon typically contains a timestamp $t_{1}$ and geographical location. This message is received by $\mathcal{B}$ at $t_{2}$ and ($t_{2} - t_{1}$) is used to verify if \ac{ToF} matches the expected distance based on the geographical proximity (computed based on the included location of $\mathcal{A}$ and own location).

The \textit{CR}-TL protocol has $\mathcal{A}$ send a timed challenge to $\mathcal{B}$, which must respond within a minimal delay. The \textit{CR} serves as a ranging message exchange and does not involve computationally non-negligible operations to have an accurate \ac{ToF} measurement (with a low constant for responding). The \ac{ToF} and geographical distances are expected to be practically the same. The underlying requirement is that $\mathcal{A}$ and $\mathcal{B}$ have trustworthy location data, while \ac{LoS} is necessary for availability. 

An \ac{SND} protocol is considered complete if the following properties are satisfied~\cite{poturalski2013formal}:

\noindent\textbf{P1 -- Correctness} ensures that if an honest device (node) is declared neighbor at some time $t$, the node must indeed be a communication neighbor at that time ($\mathcal{A}$ being able to receive messages from $\mathcal{B}$).

\noindent\textbf{P2 -- Availability} implies that the protocol ascertains neighbor relationships for every distance within a specified \ac{ND} range ($R$), i.e., $\mathcal{R}_{\text{SND}}$ is $\leq R$, the radio/datalink nominal range.

\usetikzlibrary{decorations.pathmorphing}
\tikzset{every picture/.style={line width=0.75pt}}
\begin{figure*}[!htbp]
\centering
\begin{minipage}[b]{0.5\textwidth}
    \centering
    \tikzset{every picture/.style={line width=0.75pt}}       
    \begin{tikzpicture}[x=0.75pt,y=0.75pt,yscale=-1,xscale=1]
        \draw (94.25,111.03) node [anchor=north west][inner sep=0.75pt]   [align=left] {$\mathcal{A}$};
        \draw (235,111.66) node [anchor=north west][inner sep=0.75pt]   [align=left] {$\mathcal{B}$};
        \draw (162.39,176.55) node [anchor=north west][inner sep=0.75pt]   [align=left] {\color{red} $\mathcal{E}$};
        \draw    (113.25,122.58) -- (232,123.11) ;
        \draw [shift={(234,123.12)}, rotate = 180.26] [color={rgb, 255:red, 0; green, 0; blue, 0 }  ][line width=0.75]    (10.93,-3.29) .. controls (6.95,-1.4) and (3.31,-0.3) .. (0,0) .. controls (3.31,0.3) and (6.95,1.4) .. (10.93,3.29)   ;
        \draw [shift={(111.25,122.57)}, rotate = 0.26] [color={rgb, 255:red, 0; green, 0; blue, 0 }  ][line width=0.75]    (10.93,-3.29) .. controls (6.95,-1.4) and (3.31,-0.3) .. (0,0) .. controls (3.31,0.3) and (6.95,1.4) .. (10.93,3.29)   ;
        \draw [color={rgb, 255:red, 250; green, 0; blue, 0 }  ,draw opacity=1 ]   (183.86,177.12) -- (189.77,171.73) .. controls (189.88,169.38) and (191.12,168.25) .. (193.47,168.36) .. controls (195.82,168.47) and (197.05,167.34) .. (197.16,164.99) .. controls (197.27,162.64) and (198.5,161.51) .. (200.85,161.62) .. controls (203.2,161.73) and (204.44,160.6) .. (204.55,158.25) .. controls (204.66,155.9) and (205.89,154.77) .. (208.24,154.88) .. controls (210.59,154.99) and (211.82,153.86) .. (211.93,151.51) .. controls (212.04,149.16) and (213.28,148.03) .. (215.63,148.14) .. controls (217.98,148.25) and (219.21,147.12) .. (219.32,144.77) .. controls (219.43,142.42) and (220.66,141.29) .. (223.01,141.4) -- (226.61,138.11) -- (232.52,132.72) ;
        \draw [shift={(234,131.37)}, rotate = 137.62] [color={rgb, 255:red, 250; green, 0; blue, 0 }  ,draw opacity=1 ][line width=0.75]    (10.93,-3.29) .. controls (6.95,-1.4) and (3.31,-0.3) .. (0,0) .. controls (3.31,0.3) and (6.95,1.4) .. (10.93,3.29)   ;
        \draw [shift={(182.39,178.47)}, rotate = 317.62] [color={rgb, 255:red, 250; green, 0; blue, 0 }  ,draw opacity=1 ][line width=0.75]    (10.93,-3.29) .. controls (6.95,-1.4) and (3.31,-0.3) .. (0,0) .. controls (3.31,0.3) and (6.95,1.4) .. (10.93,3.29)   ;
        \draw [color={rgb, 255:red, 255; green, 0; blue, 0 }  ,draw opacity=1 ]   (159.93,176.8) -- (154.1,171.32) .. controls (151.75,171.39) and (150.53,170.24) .. (150.46,167.89) .. controls (150.39,165.54) and (149.17,164.4) .. (146.82,164.47) .. controls (144.47,164.54) and (143.25,163.39) .. (143.18,161.04) .. controls (143.11,158.69) and (141.89,157.54) .. (139.54,157.61) .. controls (137.19,157.68) and (135.97,156.54) .. (135.89,154.19) .. controls (135.82,151.84) and (134.6,150.69) .. (132.25,150.76) .. controls (129.9,150.83) and (128.68,149.68) .. (128.61,147.33) .. controls (128.54,144.98) and (127.32,143.84) .. (124.97,143.91) .. controls (122.62,143.98) and (121.4,142.83) .. (121.33,140.48) -- (118.53,137.85) -- (112.71,132.37) ;
        \draw [shift={(111.25,131)}, rotate = 43.25] [color={rgb, 255:red, 255; green, 0; blue, 0 }  ,draw opacity=1 ][line width=0.75]    (10.93,-3.29) .. controls (6.95,-1.4) and (3.31,-0.3) .. (0,0) .. controls (3.31,0.3) and (6.95,1.4) .. (10.93,3.29)   ;
        \draw [shift={(161.39,178.17)}, rotate = 223.25] [color={rgb, 255:red, 255; green, 0; blue, 0 }  ,draw opacity=1 ][line width=0.75]    (10.93,-3.29) .. controls (6.95,-1.4) and (3.31,-0.3) .. (0,0) .. controls (3.31,0.3) and (6.95,1.4) .. (10.93,3.29)   ;
    \end{tikzpicture}\\
\subcaption{}
\label{subfig:extended_internal_adv}
\end{minipage}%
\begin{minipage}[b]{0.5\textwidth}
    \centering
    \tikzset{every picture/.style={line width=0.75pt}} 
    \begin{tikzpicture}[x=0.75pt,y=0.75pt,yscale=-1,xscale=1]
    \draw (326.75,108.03) node [anchor=north west][inner sep=0.75pt]   [align=left] {$\mathcal{A}$};
    \draw (394.89,174.55) node [anchor=north west][inner sep=0.75pt]   [align=left] {\color{red} $\mathcal{E}_{1}$};
    \draw (396.75,108.03) node [anchor=north west][inner sep=0.75pt]   [align=left] {$\mathcal{B}$};
    \draw (464.89,174.55) node [anchor=north west][inner sep=0.75pt]   [align=left] {\color{red} $\mathcal{E}_{2}$};
    \draw [color={rgb, 255:red, 255; green, 0; blue, 0 }  ,draw opacity=1 ]   (392.44,174.64) -- (386.65,169.11) .. controls (384.3,169.16) and (383.09,168.01) .. (383.04,165.66) .. controls (382.99,163.3) and (381.78,162.15) .. (379.42,162.2) .. controls (377.07,162.25) and (375.86,161.1) .. (375.81,158.75) .. controls (375.75,156.4) and (374.54,155.25) .. (372.19,155.3) .. controls (369.84,155.35) and (368.63,154.19) .. (368.58,151.84) .. controls (368.52,149.49) and (367.31,148.34) .. (364.96,148.39) .. controls (362.61,148.44) and (361.4,147.29) .. (361.35,144.94) .. controls (361.3,142.58) and (360.09,141.43) .. (357.73,141.48) .. controls (355.38,141.53) and (354.17,140.38) .. (354.11,138.03) -- (350.98,135.03) -- (345.19,129.51) ;
    \draw [shift={(343.75,128.13)}, rotate = 43.69] [color={rgb, 255:red, 255; green, 0; blue, 0 }  ,draw opacity=1 ][line width=0.75]    (10.93,-3.29) .. controls (6.95,-1.4) and (3.31,-0.3) .. (0,0) .. controls (3.31,0.3) and (6.95,1.4) .. (10.93,3.29)   ;
    \draw [shift={(393.89,176.02)}, rotate = 223.69] [color={rgb, 255:red, 255; green, 0; blue, 0 }  ,draw opacity=1 ][line width=0.75]    (10.93,-3.29) .. controls (6.95,-1.4) and (3.31,-0.3) .. (0,0) .. controls (3.31,0.3) and (6.95,1.4) .. (10.93,3.29)   ;
    \draw [color={rgb, 255:red, 255; green, 0; blue, 0 }  ,draw opacity=1 ]   (462.44,174.64) -- (456.65,169.11) .. controls (454.3,169.16) and (453.09,168.01) .. (453.04,165.66) .. controls (452.99,163.3) and (451.78,162.15) .. (449.42,162.2) .. controls (447.07,162.25) and (445.86,161.1) .. (445.81,158.75) .. controls (445.75,156.4) and (444.54,155.25) .. (442.19,155.3) .. controls (439.84,155.35) and (438.63,154.19) .. (438.58,151.84) .. controls (438.52,149.49) and (437.31,148.34) .. (434.96,148.39) .. controls (432.61,148.44) and (431.4,147.29) .. (431.35,144.94) .. controls (431.3,142.58) and (430.09,141.43) .. (427.73,141.48) .. controls (425.38,141.53) and (424.17,140.38) .. (424.11,138.03) -- (420.98,135.03) -- (415.19,129.51) ;
    \draw [shift={(413.75,128.13)}, rotate = 43.69] [color={rgb, 255:red, 255; green, 0; blue, 0 }  ,draw opacity=1 ][line width=0.75]    (10.93,-3.29) .. controls (6.95,-1.4) and (3.31,-0.3) .. (0,0) .. controls (3.31,0.3) and (6.95,1.4) .. (10.93,3.29)   ;
    \draw [shift={(463.89,176.02)}, rotate = 223.69] [color={rgb, 255:red, 255; green, 0; blue, 0 }  ,draw opacity=1 ][line width=0.75]    (10.93,-3.29) .. controls (6.95,-1.4) and (3.31,-0.3) .. (0,0) .. controls (3.31,0.3) and (6.95,1.4) .. (10.93,3.29)   ;
    \end{tikzpicture}\\
\subcaption{}
\label{subfig:extended_external_adv}
\end{minipage}%
\caption{Extended Adversarial Model. (a) External or \textit{honest-but-curious} (internal) adversary eavesdropping communications between $\mathcal{A}$ and $\mathcal{B}$ while they execute \ac{SND} protocol. $\mathcal{E}$ eavesdrops messages exchanged to learn about $\mathcal{A}$ and $\mathcal{B}$. (b) External adversary $\mathcal{E}_1$ attempting to initiate the \ac{SND} protocol with $\mathcal{A}$ or \textit{honest-but-curious} $\mathcal{E}_2$ executing the \ac{SND} protocol at a high rate, both with the intent to learn (identity and location) about or track neighboring nodes.}
\label{fig:adv}
\end{figure*}
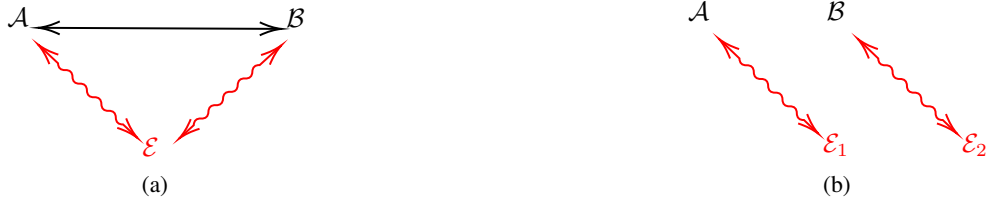
\section{Adversarial Model}
\label{sec:adversarial}

We consider both internal and external adversaries, sophisticated yet with realistic adversarial capabilities. The adversary is fundamentally a relay, $\mathcal{E}$, with one or two devices, as described in Sec.~\ref{sec:tp_snd}, that attacks \ac{SND} protocols executed by two honest nodes. We extend this adversarial model, depicted in Fig.~\ref{fig:adv}, considering first external adversaries (that is, nodes that are not provisioned with cryptographic credentials) that seek to extract/infer information about other nodes.

\begin{enumerate}
    \item Eavesdroppers, logging \ac{SND} messages to collect information on neighbor relationships node identities (for all \ac{SND} protocols) and node locations (for \textit{TL} \ac{SND} protocols). 
    \item Honest-but-curious nodes that actively initiate the SND protocol to either measure their distance to other nodes (in case of CR protocols) or have them reveal their location (for \textit{CR-TL} protocols). 
\end{enumerate}

We also consider external adversaries targeting the protocol availability by: 

\begin{enumerate}
    \setcounter{enumi}{2}
    \item Injecting forged messages for \ac{ToF} measurements (for \textit{CR} protocols) 
    \item Initiating the \ac{SND} protocol with broadcast messages to engage nearby nodes and exhaust their resources (computation, power, bandwidth).
\end{enumerate}

Then, we depart from the two-party \ac{SND} model, considering \textit{honest-but-curious} internal nodes, i.e., provisioned with cryptographic credentials and entitled to execute \ac{SND}. They execute the \ac{SND} protocol correctly but: 

\begin{enumerate}
    \setcounter{enumi}{4}
    \item They eavesdrop on \ac{SND} protocol executions by other nodes, similar to external eavesdroppers.
    \item They initiate the \ac{SND} protocol, as authorized to do, but at a high rate, with the intent to complete it, establish neighbor relationships repeatedly, and maintain fine-grained distance (\textit{CR-T} protocols) and location (\textit{CR-TL} protocols) data for their peers.
\end{enumerate}

For \textit{B} protocols, adversaries face inherent limitations due to the design of the protocol itself. As these protocols involve sending beacon messages at pre-determined, often rapid intervals, attackers cannot force \ac{SND} nodes to act more quickly/frequently than they are meant to. 

When \textit{B} protocols are executed, an eavesdropper faces significant challenges in determining who is a neighbor of whom. As beacons are typically broadcasted and do not require responses from receiving nodes, an eavesdropper cannot easily determine whether any two nodes are actually communicating directly or merely listening to the same broadcasts.

The rate of execution of \ac{SND} protocols can be either a parameter selectable by the protocol designers or dynamically decided by the nodes based on the network's state and requirements. This flexibility enables the network to adapt to varying conditions, optimizing both performance and security. However, it is essential to find the optimal \ac{SND} rate with the potential risk of information disclosure: a higher \ac{SND} rate, while providing more frequent updates and potentially higher security, also means that more information about node presence and proximity is broadcasted more often, possibly leveraged by an eavesdropper. If the protocol definition includes a minimum period $\tau_{\text{SND}}$ for \ac{SND} execution, this sets a boundary that nodes, especially \textit{honest-but-curious} ones, will not exceed. Adhering to $\tau_{\text{SND}}$ prevents nodes from initiating \ac{SND} processes (too) frequently, thus not only conserving network resources but also minimizing exposure of network topology (or node interactions). The ability of the adversary to relay fast is important - if the relay is fast enough, $T$ protocols can be defeated. On the other hand, the lower the communication range for the SND nodes, the harder it is to have adversaries with technical capabilities to perform attacks successfully. The so-called distance-decreasing attacks~\cite{clulow2006so} at the physical layer are not within the scope of this work.
\section{Problem Statement}
\label{sec:problem_statement}
While we aim for \ac{SND}, as presented in Sec.~\ref{sec:tp_snd}, we extend the adversarial model to consider adversaries that target \ac{SND} and seek to exfiltrate node information based on the \ac{SND} execution. Hence, we build on top of the proven secure \ac{SND} protocol, with the aim to design a protocol that enables secure and privacy-preserving \ac{SND}, mitigating the risk posed by \textit{honest-but-curious} internal adversaries, and external adversaries (as defined in Sec.~\ref{sec:adversarial}). Our objective is to have a protocol that thwarts learning about the identity of the nodes participating in the protocol execution and their locations, effectively preventing internal and external adversaries from being able to identify, extract, or link information eavesdropped during the protocol execution by any of the nodes.

\noindent\textbf{Properties of a Privacy-Preserving Secure Neighbor Discovery.} In addition to \textbf{P1} and \textbf{P2} (discussed in Sec.~\ref{sec:tp_snd}), an \ac{SND} protocol is considered privacy-preserving if the following properties are also met:

\noindent\textbf{P3 -- Pseudonymity\footnote{Pseudonymity can be augmented to anonymity (always along with security, notably authenticity, and integrity) with the use of anonymous authentication in schemes such as in~\cite{khodaei2019scaling}. This would require a different protocol design in our context, and it is beyond the scope of this work.}} (or pseudonymous authentication) refers to the ability of participating nodes to hide their true identities/credentials during the communication process from external adversaries, \textit{honest-but-curious} participants, and honest n. des (neighbors), while still having their messages authenticated. In particular, each node uses a typical ephemeral identity and credential, termed pseudonym, and a corresponding ephemeral private key, that cannot be linked/connected to the actual identity and long-term credential and private key of the node.

\noindent\textbf{P4 -- Confidentiality} refers to the protection of sensitive information, notably the location of nodes executing the SND protocol. External nodes cannot obtain it, and it is not revealed to the other protocol participants.

\noindent\textbf{P5 -- Unlinkability} refers to the ability of a device to participate in the protocol without its identity linked across multiple protocol executions. It should be impossible for any observer or another participating device to determine whether two \ac{SND} executions by the same node were indeed so.
\section{Proposed Solution}
\label{sec:protocol_design}
The \ac{PP-SND} protocol relies on pseudonyms to conceal the identity of the nodes participating in the protocol execution and provides authentication. Additionally, distance computations employ privacy-preserving techniques to preserve location information. This section presents the considered setup and overall design (Sec.~\ref{subsec:csdn}), and the core elements, namely, pseudonymous authentication (Sec.~\ref{subsec:pa}), privacy-preserving distance estimation (Sec.~\ref{subsec:pp-de}), of the \ac{PP-SND} protocol. We then present the design in a nutshell (Sec.~\ref{subsec:din}) and the protocol operation and its phases (Sec.~\ref{subsec:pof}).

\subsection{Considered Setup and Design Choices}
\label{subsec:csdn}
\textbf{Considered Setup.} We consider two nodes (devices), $\mathcal{A}$ and $\mathcal{B}$, which could be deployed as a part of any general infrastructure where secure communication is essential. $\mathcal{A}$ and $\mathcal{B}$ are equipped with radios that support a nominal communication range $\mathcal{R}$ and the \ac{SND} range, $\mathcal{R}_{SND} < \mathcal{R}$, a more conservative range (device distance) for which \ac{SND} is executed, and concludes with respect to $\mathcal{R}_{SND}$ whether $\mathcal{A}$ and $\mathcal{B}$ are neighbors. While $\mathcal{A}$ and $\mathcal{B}$ can be located at specific $($\text{latitude}, \text{longitude}$)$ coordinates, i.e., stationary, our setup acknowledges the possibility of mobility. Assuming that mobility does not change significantly within the protocol execution, $\mathcal{R}_{SND}$ can be chosen so that the maximum relative node speed does not cause a displacement close to $\mathcal{R} - \mathcal{R}_{\text{SND}}$.

\textbf{SND Protocol Choice.} \textit{CR}-SND protocols require an exchange of a challenge and a response between the participating nodes, inherently leading to solutions that provide mutual authentication and \ac{SND} since both parties actively participate. In this work, we design the \ac{PP-SND} protocol based on the \textit{CR}-TL \ac{SND} protocol. It necessitates an exchange between nodes, naturally utilizing both \ac{ToF} and precise geographical location coordinates. \textit{CR}-TL protocols are particularly effective against sophisticated relay attacks.

\subsection{Pseudonymous Authentication}
\label{subsec:pa}
\acp{LTC} and \ac{LTCA} are essential for accountability within a \ac{PKI}. The \ac{LTCA} is responsible for registering entities, such as devices or users, and issuing one \ac{LTC}. Using their \ac{LTC} and their corresponding long-term private key, devices can obtain anonymous tokens from the \ac{LTCA}, using, in turn, the tokens to anonymously request pseudonyms. Pseudonyms (PNYM) are ephemeral certificates \cite{jin2015scaling, khodaei2019scaling}, from a \ac{PCA}. The pseudonyms are anonymized, unique user/entity representations, replacing the \ac{LTC}, in the form of $\langle$Pseudonym Provider ID, Pseudonym Lifetime, Public Key ($pk$), Pseudonym Provider Signature$\rangle$; the PNYM Provider ID refers to the \ac{PCA} whose signature is used to validate the pseudonym and establish trust. \textit{Pseudonym Lifetime} indicates how long the pseudonym is valid. \textit{Public Key} ($pk$) together with the corresponding private key are generated by the device, with $pk$ sent via a certificate signing request to the \ac{PCA}. Similar to the \ac{IoV} \ac{PCA} and pseudonymous authentication, each node gets $K$ pseudonyms, each valid for a specific lifetime $\tau$~\cite{khodaei2020cooperative}. 

The \ac{PP-SND} protocol relies on the use of a modified pseudonym structure that includes two additional elements, namely, the Paillier public key~\cite{Paillier1999public} addition to the aforementioned public key. Given the design in~\cite{khodaei2019scaling, jin2015scaling}, and for clarity, we refer to this as the ECDSA public key. The new pseudonym becomes $\langle$Pseudonym Provider ID, Pseudonym Lifetime, Paillier Public Key ($ppk$), ECDSA public Key, Pseudonym Provider Signature$\rangle$.

\subsection{Privacy-Preserving Distance Estimation}
\label{subsec:pp-de}
\textbf{Coordinate-based Distance Estimation} can be performed using homomorphic computations on the node encrypted coordinates~\cite{zhong2007louis, hallgren2015innercircle}. This ensures that no participating nodes executing the protocol learn about one another's coordinates. A notable implementation in~\cite{nicholas_2018} utilizes \ac{HE} to estimate if an entity lies within a ``Geofence''. This is done by leveraging the \textit{Paillier} cryptosystem~\cite{Paillier1999public} homomorphic properties: it provides \ac{PHE}, allowing for the additive combination of encrypted values. Specifically, given two plaintexts, $m_1$ and $m_2$, their respective encryptions, $Enc(m_1)$ and $Enc(m_2)$, can be multiplicatively combined to yield $Enc(m_1 + m_2)$. Decryption of this result reveals the sum of $m_1 + m_2$. Furthermore, Paillier \ac{PHE} facilitates subtraction by computing the modular inverse of $Enc(m_2)$ and then multiplying it with $Enc(m_1)$, thereby obtaining $Enc(m_1 - m_2)$. Additionally, scalar multiplication is achievable, allowing for the calculation of $Enc(m_1 \cdot m_2)$ through the exponentiation of $Enc(m_1)^{m_2}$.

Initially, for the \ac{PP-SND} protocol, $\mathcal{B}$ encrypts its coordinates using $\mathcal{A}$'s $ppk$. It performs \ac{HE} operations (mainly subtraction) between the encrypted $\mathcal{A}$ $lat$ and $lng$ and its own. This allows only $\mathcal{A}$ to decrypt the difference using its $psk$ and compute the Euclidean distance, in addition to the distance based on \ac{ToF}$_{\mathcal{AB}}$, to verify if it is within range or not. This step is essential for maintaining privacy, as neither party reveals their actual coordinates.

\noindent\textbf{Time of Flight-based Distance Estimation} computes the \ac{ToF}, with $\mathcal{A}$ initiating a local timer at $t_{1}$ once it starts the discovery process. Upon the reception of the response packet from node $\mathcal{B}$ at time $t_{2}$, the timer stops. The time difference, $\delta = t_{2} - t_{1}$, is used to compute $d_{\text{ToF}} = \frac{c \cdot \delta }{2}$, where  $c$ ($3\times 10^8$) is multiplied by $\delta$, the \ac{RTT}, divided by 2 to estimate the distance.

\subsection{Design in Nutshell}
\label{subsec:din}
The \ac{PP-SND} protocol enables secure and privacy-preserving neighbor discovery in wireless networks. Each entity generates Paillier and \ac{ECDSA} key pairs for \ac{HE} operations and signing/verification, respectively. Initially, entity $\mathcal{A}$ broadcasts an advertisement message containing its identifier, an authentication token, and a pseudonym to nearby devices, maintaining pseudonymity and integrity. 

During the ranging phase, $\mathcal{A}$ sends a message to $\mathcal{B}$ with its identifier and a nonce, initiating a timer to calculate the \ac{ToF}. $\mathcal{B}$ responds with its nonce, upon receipt $\mathcal{A}$ stops its timer, allowing $\mathcal{A}$ to estimate the distance based on \ac{ToF}. To authenticate the ranging exchange, $\mathcal{B}$ sends a message back to $\mathcal{A}$, an authenticator of the exchange, also attaching its pseudonym. $\mathcal{A}$ verifies the authenticator and checks if the $d_{\text{ToF}}(\mathcal{A, B})$ is within a predefined secure range, $\mathcal{R}_{\text{SND}}$. 

In the coordinate exchange and distance calculation step, $\mathcal{A}$ encrypts its coordinates using its Paillier public key, $ppk$, and sends them to $\mathcal{B}$ along with a signed message. $\mathcal{B}$ uses $\mathcal{A}$’s $ppk$ to encrypt its coordinates and calculate the \ac{HE} difference, it then sends the encrypted coordinate difference to $\mathcal{A}$. $\mathcal{A}$ decrypts using its Paillier secret key, $psk$, the differences and computes the Euclidean distance, essentially $d_{\text{loc}}(\mathcal{A},\mathcal{B})$ without having any access to loc($\mathcal{B}$). Finally, $\mathcal{A}$ declares $\mathcal{B}$ as a neighbor if \ac{ToF} and \ac{HE}-based distances are within an acceptable low threshold, i.e., $< \epsilon$.

Combining the aforementioned components results in a secure and privacy-preserving \ac{ND}. Indeed, \ac{ToF} measurements and homomorphically encrypted location coordinates ensure that distance is derived without exposing the actual location information. Moreover, the distance between the communicating entities does not provide precise information on where the devices are, not even direction.

\subsection{Protocol Operation Phases}
\label{subsec:pof}
\begin{figure}[!h]
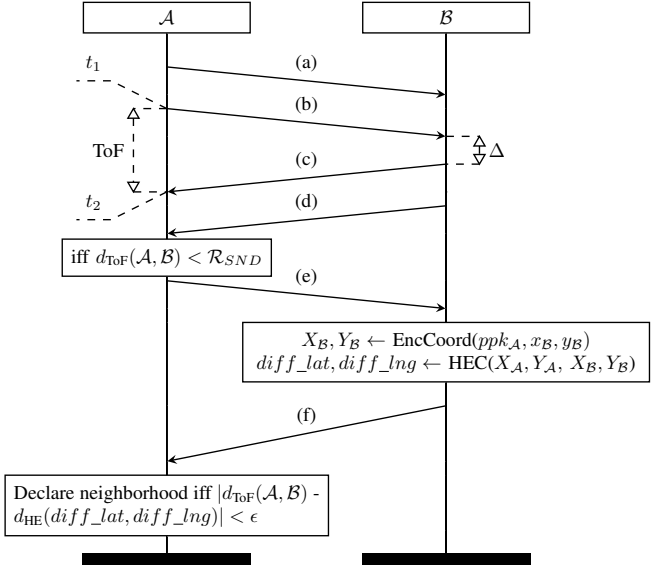

\setlength{\envinstdist}{0cm}
    \centering
    \setmsckeyword{}
    \resizebox{\columnwidth}{!}{%
    \begin{msc}[l]{}
        \setlength{\envinstdist}{2\envinstdist}
        \setlength{\instwidth}{3\mscunit}
        \setlength{\instdist}{1\mscunit} 
        \setlength{\instdist}{2cm}

        \declinst{A}{}{$\mathcal{A}$}
        \declinst{B}{}{$\mathcal{B}$}
        \mess{(a)}{A}{B}[1]
        \nextlevel[1.5]
        \mess{(b)}{A}{B}[1]
        \mscmark{$t_{1}$}{A}
        \measure[side=left]{ToF}{A}{A}[3]
        \nextlevel[1]
        \measure[side=right]{$\Delta$}{B}{B}[1]
        \nextlevel[1]
        \mess{(c)}{B}{A}[1]
        \nextlevel[1]
        \mscmark[position=below]{$t_{2}$}{A}
        \nextlevel[0.5]
        \mess{(d)}{B}{A}[1]
        \nextlevel[1.25]
        \action*{iff $d_{\text{ToF}}(\mathcal{A}, \mathcal{B}) < \mathcal{R}_{SND}$}{A}
        \nextlevel[1.5]
        \mess{(e)}{A}{B}[1]
        \nextlevel[1.5]
        \action*{\parbox{7cm}{\centering $X_{\mathcal{B}}, Y_{\mathcal{B}} \leftarrow$ EncCoord($ppk_{\mathcal{A}}, x_{\mathcal{B}}, y_{\mathcal{B}})$ \\ $diff\_lat, diff\_lng \leftarrow$ HEC($X_{\mathcal{A}}, Y_{\mathcal{A}},$ $X_{\mathcal{B}}, Y_{\mathcal{B}}$)}}{B}
        \nextlevel[3]
        \mess{(f)}{B}{A}[2]
        \nextlevel[2.5]
        \action*{\parbox{5.5cm}{Declare neighborhood iff $|d_{\text{ToF}}(\mathcal{A}, \mathcal{B})$ - $d_{\text{HE}}(diff\_lat, diff\_lng)| < \epsilon$}}{A}
        \nextlevel[2]
        \end{msc}%
    }
    \caption{\ac{PP-SND} Protocol.}
    \label{fig:protocolArch}
\end{figure}

Recall that each entity in the current PNYM lifetime has two sets of keys before engaging in the protocol: Paillier key pair ($ppk,~psk$) for \ac{HE} operations and ECDSA key pair for signing and verifying. The PNYM includes Paillier public key ($ppk$) and \ac{ECDSA} public key. Additionally, $auth_\mathcal{A}$ and $auth_\mathcal{B}$ are digital signatures computed with the node's current private key (corresponding to the current PNYM public key). Fig.~\ref{fig:protocolArch} depicts the protocol messages sequence.

First, an initialization/advertisement message is transmitted by $\mathcal{A}$, which initiates the protocol by broadcasting a message to all nearby devices. This message (a) includes its identifier $\mathcal{A}$, an authentication $auth_{\mathcal{A}}(n_{1})$, and $PNYM_{\mathcal{A}}$. This broadcast announces $\mathcal{A}$'s presence while keeping its identity hidden. Moreover, $auth_{\mathcal{A}}(n_{1})$ ensures authenticity, and integrity, proving that $\mathcal{A}$ is a legitimate participant. 

Second, $\mathcal{A}$ sends a follow-up message (b), initiating ranging, containing $\mathcal{A}$'s identifier and the nonce $n_{1}$. This step initiates a timer (at $t_{1}$) on $\mathcal{A}$, to calculate the \ac{ToF} that will be used for $d_{\text{\ac{ToF}}}(\mathcal{A}, \mathcal{B})$. $\mathcal{A}$ waits for $\mathcal{B}$'s response. $\mathcal{B}$ checks if $h(n_1)$, using $n_1$, equals the field in message (a) and aborts the protocol run otherwise. Then $\mathcal{B}$ replies (c) with a nonce $n_{2}$, and $\mathcal{A}$ stops the timer (at $t_{2}$) upon receipt, and computes $d_{\text{ToF}}$. This exchange enables $\mathcal{A}$ to estimate the distance based on \ac{ToF}. $\mathcal{B}$ sends a follow-up message (d) that contains $auth_{\mathcal{B}}(n_{1}, n_{2})$, and its $PNYM_\mathcal{B}$. This enables $\mathcal{A}$ to verify if $\mathcal{B}$ is a legitimate and honest node. $\mathcal{A}$ then checks if $d_{\text{ToF}}(\mathcal{A}, \mathcal{B}) < \mathcal{R}_{SND}$ to either continue the execution of the protocol or terminates.

Once $\mathcal{A}$ verifies the $d_{\text{ToF}}(\mathcal{A}, \mathcal{B}) < \mathcal{R}_{SND}$, it encrypts its coordinates (lat, lng) using its Paillier public key, $ppk_{\mathcal{A}}$, and sends it to $\mathcal{B}$ along with the $auth_{\mathcal{A}}(message)$ (message refers to the entire payload). This is the message (e) in the exchange. $\mathcal{B}$ then uses $ppk_{\mathcal{A}}$ to encrypt its own coordinates and perform \ac{HE} computations on the encrypted data to obtain the coordinate differences. $\mathcal{B}$ then sends a message (f) containing the encrypted differences ($diff\_lat, diff\_lng$), and $auth(message)$. $\mathcal{A}$ decrypts the difference using its own $psk_{\mathcal{A}}$ and estimate the euclidean distance. Finally, $\mathcal{B}$ is declared a neighborhood iff $|d_{\text{ToF}}(\mathcal{A}, \mathcal{B}) - d_{\text{HE}}(diff\_lat, diff\_lng)| < \epsilon$. The sequence of message exchanges is summarized as follows:

\begin{table}[!h]
    \centering
    \begin{tabularx}{\columnwidth}{lX}
        (a) & $\mathcal{A}\rightarrow * :\left\langle \mathcal{A}, h(n_{1}), auth_{\mathcal{A}}(n_{1}), PNYM_\mathcal{A} \right\rangle$  \\ \\
        
        (b) & $\mathcal{A}\rightarrow * :\left\langle \mathcal{A}, n_{1} \right\rangle$ \textit{\color{red} $\#$ Ranging message}\\ \\ 
        
        (c) & $\mathcal{B}\rightarrow \mathcal{A} :\left\langle \mathcal{B}, \mathcal{A}, n_{2} \right\rangle$ \textit{\color{red}  $\#$ Ranging message}\\ \\
        
        (d) & $\begin{array}{@{}l@{}}\mathcal{B}\rightarrow \mathcal{A} :\langle \mathcal{B}, \mathcal{A}, n_{2} + 1, auth_{\mathcal{B}}(\mathcal{A}, n_{1}, n_{2} + 1), PNYM_\mathcal{B} \rangle \end{array}$ \\ \\
        
        (e) & $\begin{array}{@{}l@{}}\mathcal{A}\rightarrow \mathcal{B} :\langle \mathcal{A}, \mathcal{B}, X_{\mathcal{A}}, Y_{\mathcal{A}}, auth_{\mathcal{A}}(X_{\mathcal{A}}, Y_{\mathcal{A}}) \rangle \end{array}$ \\ \\
        
        (f) & $\begin{array}{@{}l@{}}\mathcal{B}\rightarrow \mathcal{A} :\langle \mathcal{B}, \mathcal{A}, diff\_lat, diff\_lng,\\ 
        auth_{\mathcal{B}}(diff\_lat, diff\_lng) \rangle \end{array}$ \\ \\ 
        Final Step & $\mathcal{A}: \text{Declare neighborhood iff } |d_{\text{ToF}}(\mathcal{A}, \mathcal{B}) - d_{\text{HE}}(diff\_lat, diff\_lng)| < \epsilon$
    \end{tabularx}
\end{table}

\textbf{EncCoord} (Algorithm~\ref{algo:ec}) and \textbf{HEC} (Algorithm~\ref{algo:hec}) are used to encrypt coordinates using a given $ppk$ and perform \ac{HE} difference, respectively. \textbf{EncCoord} initially uses Algorithm~\ref{algo:norma} to normalize the input values (degree) to achieve better accuracy. 

\drawframe{no}

\begin{algorithm}[!h]
{\small
\SetAlgoLined
\KwIn{$public\_key$, $lat$, $lng$}
\KwOut{$encrypted\_lat$, $encrypted\_lng$}
$normalized\_lat \gets normalize\_deg(lat)$\;
$normalized\_lng \gets normalize\_deg(lng)$\;
$encrypted\_lat \gets public\_key.encrypt(normalized\_lat)$\;
$encrypted\_lng \gets public\_key.encrypt(normalized\_lng)$\;
\Return $(encrypted\_lat, encrypted\_lng)$\;
}
\caption{Encrypt Coordinates (EncCoord)}
\label{algo:ec}
\end{algorithm}

\begin{algorithm}[h!]
{\small
\SetAlgoLined
\KwIn{$encrypted\_input\_lat$, $encrypted\_input\_lng$, $my\_encrypted\_lat$, $my\_encrypted\_lng$}
\KwOut{$diff\_lat$, $diff\_lng$}
$diff\_lat \gets encrypted\_input\_lat - my\_encrypted\_lat$\;
$diff\_lng \gets encrypted\_input\_lng - my\_encrypted\_lng$\;
\Return $(diff\_lat, diff\_lng)$\;
}
\caption{Homomorphically Compute the Difference Between Input and Participant Coordinates (HEC)}
\label{algo:hec}
\end{algorithm}

\begin{algorithm}[!h]
\SetAlgoLined
{\small
\KwIn{$deg$, $is\_lat$}
\KwOut{$normalized\_deg$}
\If{$is\_lat$}{
    $normalized\_deg \gets deg + 90$\;
} \Else{
    $normalized\_deg \gets deg + 180$\;
}
$normalized\_deg \gets normalized\_deg \cdot normalize\_factor$\;
$normalized\_deg \gets round(normalized\_deg)$\;
\Return $normalized\_deg$\;
}
\caption{Normalize Degrees for Encryption}
\label{algo:norma}
\end{algorithm}
\section{Performance Evaluation and Security \& Privacy Analysis}
\label{sec:performanceANDanalysis}
This section provides details on \ac{SND} and \ac{PP-SND} cryptographic performance overhead evaluated on Raspberry Pi (Sec.~\ref{sec:performance}). Additionally, achieved security and privacy properties are analyzed (Sec.~\ref{sec:analysis}). 

\subsection{SND and PP-SND Performance Evaluation}
\label{sec:performance}

The protocol implementation is done using Python $3.11$. We selected the \textit{Paillier} cryptosystem~\cite{Paillier1999public}, known for its multiplicative and additive homomorphic properties (as earlier discussed in Sec.~\ref{subsec:pa}) with the use of the \textit{phe} 1.5.0 python library~\cite{PythonPaillier}. Additionally, the gmpy2~\cite{gmpy2} extension is used as it supports fast multiple-precision arithmetic to speed up the \ac{HE} computations. \ac{ECDSA} was employed for digital signatures, specifically utilizing the \textit{brainpoolP256r1} curve parameters~\cite{chen2023recommendations}. The signing and verifying keys are $256$ bits each. To assess the efficiency and performance, we benchmark the cryptographic overhead imposed through a $10,000$ simulations, i.e., emulated executions of the protocol---with the processing by the actual device. This is done using different key sizes, i.e., 1024, 2048, and 3072 bits, corresponding to different security levels, i.e., 80, 112, and 128 bits, recommended by NIST~\cite{keylength}. The execution time mean value is then computed, in addition to the 95\% \ac{CI}. We evaluated the protocol on a Raspberry Pi 4, running a 4 cores Broadcom BCM2711 Quad-core Cortex-A72 (ARM v8) CPU, with a clock speed of 1.8 GHz, and 8 GB of RAM.

\begin{figure*}[t!]
\centering
  \begin{subfigure}[b]{0.33\textwidth}
    \includegraphics[width=\textwidth]{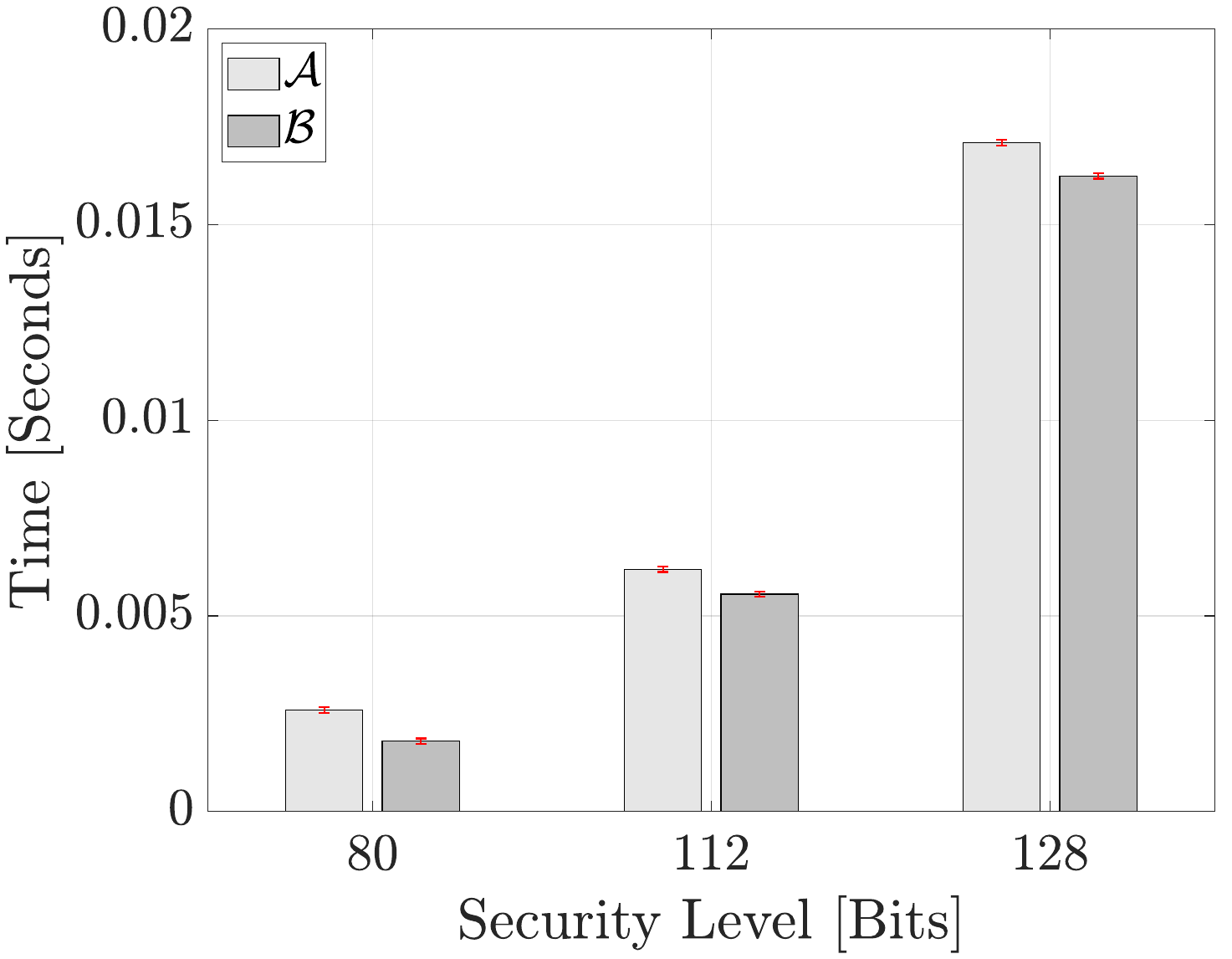}
    \caption{}
    \label{fig:SND_RBPi}
  \end{subfigure}
   \begin{subfigure}[b]{0.33\textwidth}
    \includegraphics[width=\textwidth]{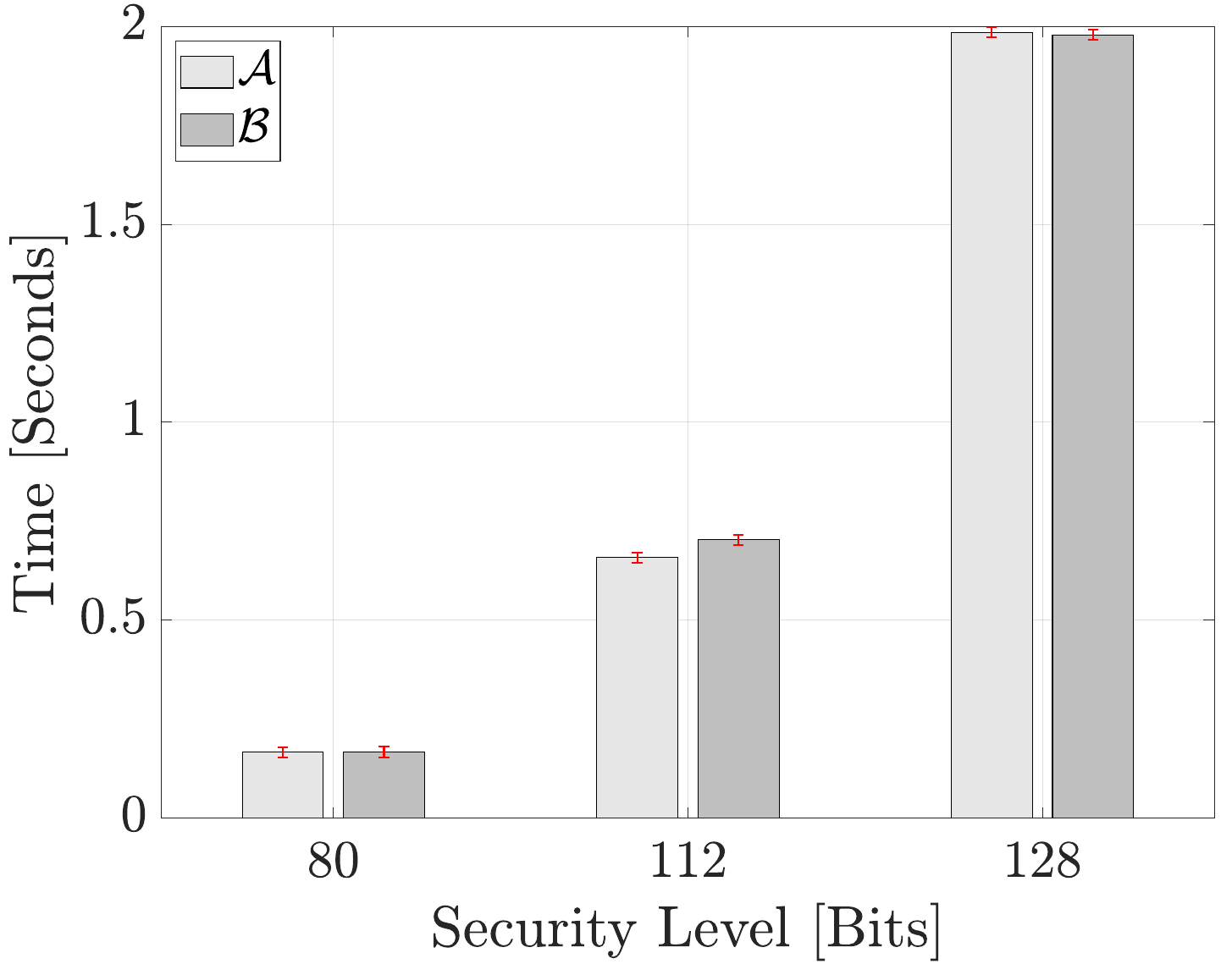}
    \caption{}
    \label{fig:PP_SND_RBPi}
  \end{subfigure}
    \caption{Evaluation of the cryptographic overhead for each protocol, (a) \ac{SND}, and (b) \ac{PP-SND}, as a function of different security levels.}
    \label{fig:all_evaluation}
\end{figure*}

Fig.~\ref{fig:all_evaluation} illustrates the overall performance of the two protocols, i.e., \ac{SND} and \ac{PP-SND}, for $\mathcal{A}$ and $\mathcal{B}$ as a function of security level. Fig.~\ref{fig:SND_RBPi} shows the \ac{SND} performance. Indeed, increasing the key size increases computation overhead and, hence, more processing time. In all executions, $\mathcal{A}$ takes more time than $\mathcal{B}$, due to verifying the authenticated message of $\mathcal{B}$ and computing the distance. 

Similarly, Fig.~\ref{fig:PP_SND_RBPi} shows the \ac{PP-SND} performance. Unlike \ac{SND}, the \ac{PP-SND} results in higher overhead, due to the time needed to perform the \ac{HE} computation, i.e., encrypting the coordinates of $\mathcal{A}$ and $\mathcal{B}$, performing the subtraction on the encrypted values, and decrypting the difference to estimate the distance. Hence, the computation overhead of $\mathcal{A}$ and $\mathcal{B}$ is similar. $\mathcal{A}$ encrypts its own coordinates using $ppk_{\mathcal{A}}$. Similarly, $\mathcal{B}$ does the same to its coordinates using $ppk_{\mathcal{A}}$ (Algorithm~\ref{algo:ec}), and then perform subtraction (Algorithm~\ref{algo:hec}). Additionally, $\mathcal{A}$ decrypts the final difference. These operations significantly increase the execution time. Furthermore, increasing the security level (thus the key size) increases processing time.

Although the \ac{PP-SND} adds significant overhead compared to the SND execution time, it ensures the protocol participants have a secure and privacy-preserving \ac{SND} without leaking any information about the involved parties. 

\subsection{Security \& Privacy Analysis}
\label{sec:analysis}

\noindent\textbf{\ac{SND} Protocol.} The proven secure \ac{SND} protocol in~\cite{poturalski2013formal} utilizes \ac{ToF} measurements and location information to prevent potential adversaries from relaying or replaying messages between nodes in the network. The presented \ac{PP-SND} protocol leverages \ac{ToF} measurements and uses \ac{HE} techniques for the coordinate-based distance estimation, ensuring that they are (within $\mathcal{R}_{\text{SND}}$) and, most important, within $\epsilon$ of each other, fulfilling \textbf{P1} and \textbf{P2}. 

\noindent\textbf{Correctness (P1)} $\mathcal{A}$ and $\mathcal{B}$ are declared neighbors if and only if the estimated distance using \ac{ToF} and $d_{\text{loc}}(\mathcal{A}, \mathcal{B})$, while $\mathcal{R}_{\text{SND}}$ is the maximum \ac{SND} range: $d_{\text{loc}}(\mathcal{A}, \mathcal{B}) \sim d_{\text{ToF}}(\mathcal{A}, \mathcal{B})$ $\leq \mathcal{R}_{\text{SND}}, \text{ then } \mathcal{A}$ $\text{ and } \mathcal{B}$ are considered neighbors. 

\noindent\textbf{Availability (P2)} is achieved if for any $d$ such that $0 < d \leq \mathcal{R}_{\text{SND}}$, there exists a mechanism that ensures: $\exists\, d_{loc}(\mathcal{A}, \mathcal{B})$ calculated using \ac{ToF} and \ac{HE} such that $\{d_{\text{ToF}}(\mathcal{A}, \mathcal{B}) \land d_{\text{HE}}(\mathcal{A}, \mathcal{B})\} \approx d_{loc}(\mathcal{A}, \mathcal{B}) $ enabling the protocol to identify a node as a neighbor within the operational range. Let $\mathcal{A}$ and $\mathcal{B}$ be two nodes participating in the protocol, with \ac{ToF} and \ac{HE} privacy-preserving distance estimations. Let $\delta_{1} \doteq |d_{\text{ToF}}(\mathcal{A}, \mathcal{B})~-~d_{loc}(\mathcal{A}, \mathcal{B})|$ be the ToF estimate error and $\delta_{2} \doteq |d_{\text{HE}}(\mathcal{A},$ $\mathcal{B})~-~d_{loc}(\mathcal{A}, \mathcal{B})|$ be the HE estimate error. Since we know that $\delta_{1}$ is negligible~\cite{poturalski2013formal}, $\delta_{1} = \delta_{2}$, then P1 and P2 hold.

\noindent\textbf{Pseudonymity and Identity Protection.} The use of pseudonyms plays a key role in hiding user identities. By substituting real identifiers with temporary, dynamically changing pseudonyms, the \ahmedupdate{\ac{PP-SND}} protocol hides the real identity of the communicating parties. This approach effectively prevents adversaries from linking these pseudonyms to the actual user. Hence, satisfying both properties \textbf{P3} and \textbf{P5}.

Let $\mathcal{N} = {\mathcal{A}, \mathcal{B}, \ldots}$ denote the set of nodes participating in the protocol. Let $\mathcal{P} = {P_{\mathcal{A}}, P_{\mathcal{B}}, \ldots}$ denote the set of pseudonyms corresponding to the nodes in $\mathcal{N}$. For any node $\mathcal{A} \in \mathcal{N}$, $P_{\mathcal{A}}^{(t)}$ denotes the pseudonym used by $\mathcal{A}$ at time $t$. 

\noindent\textbf{Pseudonymity (P3)} is preserved since each entity uses a sequence of pseudonymous identities (certificates) that cannot be connected to its actual long-term identity (\ac{LTC}). Any observer (including adversaries or other protocol participants) cannot link different pseudonyms to the same entity, thereby preserving pseudonymity. By design, not even the issuer of the pseudonyms---the \ac{PCA}---knows which \ac{LTC} they were issued for. This ensures that the PCA cannot link a pseudonym back to the \ac{LTC} or the long-term identity of the entity. As a result, unless the node itself uses its long-term private key and \ac{LTC}, no information about the entity's identity is revealed.

\noindent\textbf{Unlinkability (P5)} is achieved by utilizing a new pseudonym at each protocol execution instance with the corresponding cryptographic keys. Any observer, protocol participant, or adversary cannot determine whether this new pseudonym belongs to the same node as any previously observed pseudonym, based on the included cryptographic keys and operations; i.e., syntactic unlinkability. Each pseudonym $P_{\mathcal{A}}^{(t)}$ is generated to be independent of the previous one $P_{\mathcal{A}}^{(t-1)}$, and all $P_{n}^{(t)}, \forall n \in \mathcal{N}$, are issued with the same validity period, thereby preventing linkability through pattern analysis or identifier correlation.

Furthermore, we consider semantic unlinkability~\cite{khodaei2020cooperative}, which extends beyond the identifiers to include the context and content of the messages exchanged. Semantic unlinkability ensures that even if the content of the messages is analyzed, an adversary cannot infer that different messages are related or originated from the same node. The content of the \ac{PP-SND} messages is encrypted (homomorphically), thus not leaving any useful information to the observer or participant to link interactions once the pseudonym and the identity of the device and its public key change (transition to the next identity/pseudonym lifetime). This could have been the case if, for example, location were visible and the adversary could infer based on mobility patterns~\cite{buttyan2009slow}. Additionally, in a static setup, the computed distance can lead the \textit{honest-but-curious} to infer that two sessions could have been executed with the same peer node.

In each session, a node can select a new pseudonym with new private keys not linked to its real identity or any previous pseudonym/keys it has used. The protocol implementation is expected to ensure that all device identifiers across the protocol stack are changed at the time of the pseudonym/identity transition. For example, this can be done in modern platforms by changing the MAC address and obtaining a new IP address.

Moreover, observing, recording, or analyzing the pseudonyms used in multiple sessions provides no useful information to link these sessions to each other or the real identity of a node. Thus, even if $\mathcal{E}$ could observe all communication sessions, the lack of any deterministic pattern or link between the pseudonyms across sessions makes it computationally impractical to achieve meaningful linkage.

\noindent\textbf{Confidentiality (P4):} The content, notably location, remains inaccessible to adversaries. The exchange of information computed via \ac{HE} ensures that an adversary cannot deduce the location of the communicating parties. \ac{HE} allows for operations to be performed on encrypted data, ensuring that the outcome remains encrypted and can be only decrypted by the intended recipient, i.e., the node that holds the corresponding $psk$ to its $ppk$. 

Let $C_{\mathcal{A}} = (lat_{\mathcal{A}},~lon_{\mathcal{A}})$ and $C_{\mathcal{B}} = (lat_{\mathcal{B}},~lon_{\mathcal{B}})$ represent the geographical coordinates of $\mathcal{A}$ and $\mathcal{B}$, respectively. $E_{ppk}(x)$ and $D_{sk}(x)$ denote the encryption and decryption of $x$ using the Paillier public key $ppk$ and the private key $psk$, respectively. $E_{ppk_{\mathcal{A}}}(CD_\mathcal{AB}) = (E_{ppk_{\mathcal{A}}}(lat_\mathcal{A})~-~E_{ppk_{\mathcal{A}}}(lat_\mathcal{B}),$ $E_{ppk_{\mathcal{A}}}(lon_\mathcal{A})$ $-~E_{ppk_{\mathcal{A}}}(lon_\mathcal{B}))$ and $E_{ppk_{\mathcal{B}}}(CD_\mathcal{BA}) = (E_{ppk_{\mathcal{B}}}(lat_\mathcal{B})~-~E_{ppk_{\mathcal{B}}}(lat_\mathcal{A}),~E_{ppk_{\mathcal{B}}}(lon_\mathcal{B})~-~E_{ppk_{\mathcal{B}}}(lon_\mathcal{A}))$ represents the homomorphically computed coordinates difference between $\mathcal{AB}$, and $\mathcal{BA}$, respectively.

The Paillier Cryptosystem supports additive homomorphic encryption, where for any two plaintexts $a$ and $b$~\cite{Paillier1999public}: $E_{ppk}(a) \times E_{ppk}(b) \equiv E_{ppk}(a + b) \mod n^2,$ where $n$ is part of the public key in the Paillier Cryptosystem, the distance function $d_{sk_{\mathcal{A}}}(CD_\mathcal{AB})$ enables $\mathcal{A}$ to decrypt the computed difference, using its Paillier $psk$, and compute the Euclidean distance between itself and $\mathcal{B}$. Similarly, $d_{sk_{\mathcal{B}}}(CD_\mathcal{BA})$ enables $\mathcal{B}$ to decrypt the computed difference, using its $psk$, and to compute the Euclidean distance between itself and $\mathcal{A}$.

By encrypting $C_{\mathcal{A}}$ and $C_{\mathcal{B}}$ with the Paillier public key of the participating entities, we ensure the encrypted coordinates, $E_{ppk}(C_{\mathcal{A}})$ and $E_{ppk}(C_{\mathcal{B}})$, are not disclosed and that arithmetic operations can be computed homomorphically without decryption.

The properties of the Paillier Cryptosystem ensure that without access to either entity's private key $psk$, it does not reveal any information about the exchanged $C$. The operation on these encrypted values, namely, distance computation, remains secure and private. The encrypted difference is decrypted, and the distance is computed, then compared with the \ac{ToF}-based distance measurement, $d_{\text{ToF}}(\mathcal{A, B})$, without revealing the actual locations: 
\begin{center}
    $D_{sk_{\mathcal{A}}}(E_{ppk_{\mathcal{A}}}(CD_\mathcal{AB})) = d_{\text{loc}}(C_\mathcal{A}, C_\mathcal{B}) \approx d_{\text{ToF}}(\mathcal{A, B})$
\end{center}
        
This approach guarantees that data will not reveal the location or movement patterns of the parties involved. \ahmedupdate{Therefore, fulfilling properties \textbf{P1}, \textbf{P3}, \textbf{P4} and \textbf{P5}}. 

\noindent\textbf{Limitations.} It is important to note that this analysis does not extend to adversaries capable of localizing devices based on physical layer features, such as \ac{RSS}, or those that fingerprint devices based on unique characteristics of their radio equipment~\cite{lin2020wireless, AlHazbiAHSSGOPP:C:2024}. Such methods require additional countermeasures beyond the scope of the \ac{PP-SND} protocol.
\section{Related Work}
\label{sec:rw}
The notion of \textbf{\acl{SND}} was first introduced by~\cite{PapadimitratosH_2003}, in the context of secure route discovery---however, that property was closer to authenticated \ac{ND}. Poturalski et al.~\cite{poturalski2008secure} explored solutions for \ac{SND}~\cite{papadimitratos2008secure} by proposing protocols robust against relay attackers. Additionally, in~\cite{poturalski2013formal}, they presented a comprehensive framework for the security assessment of \ac{ND} protocols within wireless networks. They presented formal proofs of the security properties of four \ac{SND} protocols. By integrating time-and-location information into the protocol design, they showed that \ac{SND} can be achieved against fast-relaying adversaries.

\textbf{Distance-bounding Protocols} prevent distance fraud, relevant in our context if one of the SND participants were adversarial (recall: we consider only honest but curious internal nodes). Brands and Chaum~\cite{brands1993distance} pioneered these protocols to establish practical upper bounds on physical distances between parties by measuring single-bit challenge-response delays. Hancke and Kuhn~\cite{hancke2005rfid} extended this concept to RFID systems, developing a protocol that uses \ac{UWB} pulse communication. Their low-power, asynchronous approach proves particularly effective for passive RFID tokens in noisy environments, ensuring tokens remain within verified distances from authenticators. In~\cite{rasmussen2010realization}, the authors addressed implementation challenges by developing a prototype that processes signals in less than 1~$ns$ using \ac{CRCS}. This achievement limits adversarial provers from falsely appearing closer than 15~cm to the verifier, while avoiding the demodulation delays inherent in traditional XOR and comparison methods.

\textbf{Privacy-Preserving Location Estimation.} Several studies have explored privacy-preserving location estimation using \ac{HE}. Hallgren et al.~\cite{hallgren2015innercircle} developed InnerCircle, a decentralized protocol that enables proximity detection without a trusted third party. Their approach combines \ac{SMC} and \ac{PHE} to let users verify their relative distance while keeping their exact locations private. Similarly, Zhong et al.~\cite{zhong2007louis} proposed three protocols---Louis, Lester, and Pierre---for privacy-aware proximity detection in \ac{LBS}. While Louis relies on a semi-trusted third party, Lester and Pierre operate in a fully decentralized manner. All three protocols use \ac{HE} to enable users to check their friends' proximity without exposing location data to any central authority.
\section{Conclusion and Future Work}
\label{sec:conclusion}
We presented the first Privacy-Preserving Secure Neighbour Discovery (\ac{PP-SND}) protocol, where the participating entities' information, notably real identity and location, is anonymized and encrypted, making it impossible for participating entities to learn about one another. We defined the properties required to have a \ac{PP-SND} protocol that is both functional and privacy-preserving. We presented the overall protocol architecture, functions, and operations. Our findings demonstrate the protocol feasibility to provide secure and privacy-preserving device discovery, useful in ubiquitous computing scenarios. As for future work, we plan to implement and test the protocol across various wireless technologies. Moreover, we plan to test different cryptosystems that enable homomorphic operations on data and ensure their applicability across these technologies.

\section*{Acknowledgment}
This work is supported in parts by the Swedish Research Council and the Knut and Alice Wallenberg Foundation.

\bibliographystyle{IEEEtran}
\balance
\bibliography{main}

\end{document}